# The cosmological constant problem, antimatter gravity and geometry of the Universe

Dragan Slavkov Hajdukovic[1]
PH Division CERN
CH-1211 Geneva 23
dragan.hajdukovic@cern.ch
[1]On leave from Cetinje, Montenegro

**Abstract**
This article is based on two hypotheses. The first one is the existence of the gravitational repulsion between particles and antiparticles. Consequently, virtual particle-antiparticle pairs in the quantum vacuum may be considered as gravitational dipoles. The second hypothesis is that the Universe has geometry of a four-dimensional hyper-spherical shell with thickness equal to the Compton wavelength of a pion, which is a simple generalization of the usual geometry of a 3-hypersphere. It is striking that these two hypotheses lead to a simple relation for the gravitational mass density of the vacuum, which is in very good agreement with the observed dark energy density.

1. **Introduction**

No one knows what dark energy is, but we need it to explain (the recently discovered) accelerated expansion of the Universe. The most elegant and natural solution is to identify dark energy with the energy of the quantum vacuum predicted by Quantum Field Theory(QFT) ; but the trouble is that QFT predicts the energy density of the vacuum to be orders of magnitude larger than the observed dark energy density

$$\rho_{de} \approx 7.5 \times 10^{-27} \, kg/m^3 \tag{1}$$

Summing the zero-point energies of all normal modes of some field of mass $m$ up to a wave number cut-off $K_c \gg m$, QFT yields [1] a vacuum energy density (with $\hbar = c = 1$)

$$\langle \rho_{ve} \rangle = \int_0^{K_c} \frac{4\pi k^2 dk}{(2\pi)^3} \frac{1}{2}\sqrt{k^2 + m^2} \approx \frac{K_c^4}{16\pi^2} \tag{2}$$

or reintroducing $\hbar$ and $c$ and using the corresponding mass cut-off $M_c$ instead of $K_c$

$$\rho_{ve} = \frac{1}{16\pi^2}\left(\frac{c}{\hbar}\right)^3 M_c^4 \equiv \frac{\pi}{2}\frac{M_c}{\lambda_{Mc}^3} \tag{3}$$

where $\lambda_{Mc}$ denotes the (non-reduced) Compton wavelength corresponding to $M_c$. If we take the Planck scale (i.e. the Planck mass) as a cut-off, the vacuum energy density calculated from Eq. (3) is $10^{121}$ times larger than the observed dark energy density (1). If we only worry about zero-point energies in quantum chromodynamics, Eq. (3) is still $10^{41}$ times larger than the Eq. (1). Even if the Compton wavelength of an electron is taken as cut-off, the result exceeds the observed value by nearly 30 orders of magnitude. This huge discrepancy is known as the cosmological constant problem [1]. It is very unlikely that a solution of the cosmological constant problem can be found in the framework of the major approaches: superstrings, adjustment mechanisms, changing the rules of the classical General Relativity, quantum cosmology and anthropic principle (See Review [1] for more details).

In the present article the cosmological constant problem is approached on the assumptions that the gravitational interaction between matter and antimatter is repulsive, and that the Universe has geometry of a hyper-spherical shell (embedded in a four-dimensional Euclidian space) with thickness equal to the Compton wavelength of a pion. Under these assumption the relation

$$\rho_{de} = \frac{3m_\pi}{\lambda_\pi^2 R} \tag{4}$$

between the dark energy density, the pion mass $m_\pi$, pion Compton wavelength $\lambda_\pi$, and cosmological scale factor $R$ is obtained. The later is given by the Friedman equation

$$\frac{kc^2}{H^2 R^2} = \Omega - 1; \quad k = +1, -1, 0 \tag{5}$$

with $\Omega$ the total energy density of the universe relative to the critical density; $H$ is the Hubble parameter, and we restrict to a closed universe ($k = 1$). Eq. (4) is in very good agreement with the observationally inferred value Eq. (1) for the dark energy density.

In a recent and potentially very significant publication [2] it was argued that the well known Veneziano ghost in Quantum Chromodynamics (QCD) might act as a source for the cosmological constant. While our approach and approach in reference [2] are very different, they share a common point of view that only the QCD vacuum is significant in Cosmology. A series of the striking "coincidences" involving pions [3]-[5] should be considered as a support to this point of view.

## 2. Conjecture of the gravitational repulsion between particles and antiparticles and its consequences

The gravitational properties of antimatter are still not known. Our ignorance may end with the eventual accomplishment of two emerging experiments; AEGIS [6] at CERN and AGE [7] at Fermilab; both designed to measure the gravitational acceleration of atoms of antihydrogen.

While a huge majority of physicists believes that gravitational acceleration of particles and antiparticles is the same, there is room for surprises. The biggest surprise would be if the gravitational acceleration of particles and antiparticles just differs in sign and this exciting possibility is the main assumption in this paper.

The assumption of a gravitational repulsion between matter and antimatter (antigravity) as a universal property may be written

$$m_i = m_g ; \ m_i = \overline{m}_i ; \ m_g + \overline{m}_g = 0 \tag{6}$$

Here, a bar denotes antiparticles; while indices $i$ and $g$ refer to inertial and gravitational mass. The first two relations in (7) are based on experimental evidence [8]-[9], while the third is our conjecture. It radically differs from the conventional expression $m_g - \overline{m}_g = 0$, implying (together with the Newton law of gravity) that matter and antimatter are mutually repulsive bit self-attractive.

The usual statement that we live in the Universe completely dominated by matter is not true in the case of quantum vacuum, where according to QFT, virtual matter and antimatter "appear" in equal quantities. Thus, our hypothesis must have dramatic consequences for the quantum vacuum (a world of "virtual" matter) and indirectly through it to our world of "real" matter.

First, it is immediately clear that a virtual particle-antiparticle pair is a system with zero gravitational mass and such a cancelation of gravitational masses might be important for an eventual solution of the cosmological constant problem. By the way, a similar cancellation of the opposite electric charges of particle and antiparticle in a virtual pair, leads to the zero density of the electric charge of the vacuum.

Moreover, a virtual particle-antiparticle pair may be considered as a virtual gravitational dipole with dipole moment

$$\vec{p} = m\vec{d}; \quad p \approx m\lambda = \frac{h}{c} \tag{7}$$

The vector $\vec{d}$ is directed from the antiparticle to the particle, and presents the distance between them. As the distance between particle and antiparticle is of the order of the Compton wavelength, we shall use the second of Equations (7) attributing to every virtual pair a dipole moment independent of mass. Consequently, a gravitational polarization density $\vec{P}$ (i.e. the gravitational dipole moment per unit volume) may be attributed to the quantum vacuum.

As well known, in a dielectric medium the spatial variation of the electric polarization generates a charge density $\rho_b = -\nabla \cdot \vec{P}$, known as the bound charge density. In an analogous way, the gravitational polarization of the quantum vacuum should result in a gravitational mass density of the vacuum. In the rest of the paper it would be argued that dark energy may be identified with such a gravitational mass resulting from polarization.

It is evident that under the assumption of antigravity, Eqs. (2) and (3), correspond to the inertial mass density of the physical vacuum, while the dark energy is related to the density of the gravitational mass. Eqs. (2) and (3) are not in conflict with observations; they simply determine the value of a non-relevant quantity, the inertial mass instead of the gravitational mass of the quantum vacuum.

## 3. Universe as a hyper-spherical shell

From mathematical point of view the Universe is a three-dimensional subspace (with non-Euclidean intrinsic geometry) embedded in a four-dimensional Euclidean space. To be more specific, let us denote by $x, y, z, w$ the Cartesian coordinates of the four-dimensional Euclidean space with line element

$$d\sigma^2 = dx^2 + dy^2 + dz^2 + dw^2 \tag{8}$$

Our Universe is defined as a three-dimensional subspace; in fact a 3-hypersphere

$$x^2 + y^2 + z^2 + w^2 = R^2 \tag{9}$$

where R is the radius of the 3-hypersphere.

In spherical coordinates $R, \chi, \varphi, \phi$

$$\begin{aligned} x &= R\sin\chi\sin\varphi\cos\phi \\ y &= R\sin\chi\sin\varphi\sin\phi \\ z &= R\sin\chi\cos\varphi \\ w &= R\cos\chi \end{aligned} \tag{10}$$

the line element (8) becomes

$$d\sigma^2 = dR^2 + R^2\{d\chi^2 + \sin^2\chi(d\varphi^2 + \sin^2\varphi d\phi^2)\} \tag{11}$$

The second term on the right-hand side of Eq. (11) is the Friedman-Robertson-Walker's metric, i.e. intrinsic metric of 3-hypersphere (9) in comoving coordinates; the Universe expands or contracts as $R$ increases or decreases, but the galaxies keep fixed coordinates $\chi, \varphi, \phi$. Note that in the metric (11) $R$ represents a coordinate, while in FRW metric it is not a coordinate but the cosmic scale factor. From the point of view of the 4-dimensional Euclidean space, the present day Universe is an accelerated growing 3-hypersphere.

Instead of considering the Universe being a 3-hypersphere (9), assume that it is a hyper-spherical shell of small thickness $R_{ed}$. While a 3-hypersphere is a three-dimensional subspace, a hyper-spherical shell is a four dimensional subspace (of the space (11)), with one extra dimension of size $R_{ed}$. This conjecture is motivated by the expectation that vacuum fluctuations should be related to the expansion of the Universe and can't be confined on a 3-hypersphere. In other words, the gravitational dipole moment of a virtual particle-antiparticle pair should have (without violation of the cosmological principle) a nonzero component perpendicular to 3-hypersphere.

If the extra spatial dimensions exist, they must be of very small size; roughly speaking less than the size of an atom. Otherwise it would be in conflict with observation; a large extra dimension would violate the firmly established inverse square law of gravitational and electromagnetic forces.

In the present paper, we put forward the idea that there is a universal size of finite extra dimensions, related to the quantum vacuum fluctuations. More precisely, the conjecture is that an extra dimension must be sufficiently large to allow for the appearance of virtual quark-antiquark pairs, which are an inherent part of the physical vacuum in quantum chromodynamics. To our best knowledge this is the first time that the size of extra dimensions is related to vacuum fluctuations, i.e. demand that extra dimensions are not "sterile" but allow the existence of major fluctuations such as the quark-antiquark pairs. Hence, the reduced Compton wavelength of a pion should be considered as a lower-bound for the size of the extra spatial dimension. The best guess for an upper-bound should be the Compton wavelength of an electron, the lightest particle having sufficiently small Compton wavelength. (The Compton wavelength of a neutrino is of the order of one micron, what is already a size in high risk of conflict with observations).

Pions have been involved in a number of numerical "coincidences" [3]-[5], suggesting that pions may possess some (hidden) importance for the Universe as a whole. If so, the Compton wavelength of a pion should be the best candidate for a universal size of all extra dimensions. Hence, as for an approximation we may use:

$$R_{ed} \sim \lambda_\pi \tag{12}$$

At the end of this section let us remember that in a four-dimensional space with line element (12), the volume of a four dimensional ball, the volume of a hyper-spherical shell and the volume of a 3-hypersphere are respectively:

$$V_B = \frac{\pi^2}{2} R^4; \quad V_{hshell} = 2\pi^2 R^3 R_{ed}; \quad V_{hs} = 2\pi^2 R^3 \tag{13}$$

## 4. Dark Energy as a result of vacuum polarization

In the framework of the hyper-spherical shell geometry, every dipole moment has a component perpendicular to the 3-hypersphere. Assume that the four-dimensional gravitational polarization density $\vec{P}$ has a constant radial component $P_R$. Then, according to the Eq. (11) and $\rho_{ve} = -\nabla \cdot \vec{P}$, a virtual gravitational mass density

$$\rho_{v,hshell} = -\nabla \cdot P = -\frac{1}{R^3} \frac{\partial}{\partial R}\left(R^3 P_R\right) = -\frac{3P_R}{R} \tag{14}$$

may be attributed to the physical vacuum inside the hyper-spherical shell. Hence, from equations (13) and (14), the total gravitational mass of the vacuum in the hyper-spherical shell is

$$M_{v,hshell} = \rho_{v,hshell} V_{hshell} = -6\pi^2 R^2 R_{ed} P_R \qquad (15)$$

Dividing Eq. (15) by the volume of the 3-hypersphere, yields to the familiar three-dimensional gravitational mass density

$$\rho_{v,hs} = -\frac{3R_{ed}}{R} P_R \qquad (16)$$

Equation (16) can be trivially transformed into Eq. (4) if $R_{ed} = \lambda_\pi$ and if the number density ($N_0$) of the gravitational dipoles is equal to the number density of virtual pions (i.e. $N_0 = 1/\lambda_\pi^3$), except for the difference in sign.

In standard Cosmology, the gravitational mass of the vacuum is supposed to be positive and according to the cosmological field equation (*n* denotes different components of the cosmological fluid)

$$\ddot{R} = -\frac{4\pi G}{3} R \sum_n \left( \rho_n + \frac{3p_n}{c^2} \right) \qquad (17)$$

the accelerated expansion is due to the negative pressure of dark energy. In the standard case, the dark energy obeys the equation of state $p_\Lambda = -\rho_\Lambda c^2$. The sum in brackets (effective gravitational mass) is negative with value $-2\rho_\Lambda$. Our case directly generates a negative gravitational mass which drives the accelerated expansion.

The geometry of a hyper-spherical shell of thickness $L_{ed} \approx \lambda_\pi$ could be the explanation why the virtual dipoles, with the Compton wavelength much shorter than the Compton wavelength of a pion, do not contribute to the gravitational mass of the quantum vacuum. In a Universe with the geometry of a 3-hypersphere, the gravitational dipoles would be aligned in a way similar to the ground state of an antiferromagnet and hence their gravitational polarization density would be zero. Adding an extra spatial dimension (i.e. the geometry of a hyper-spherical shell) has different effects on dipoles of different size. For dipoles of Compton wavelength $\lambda_m \ll \lambda_\pi$, the antiferromagnetic like order would be preserved. Only at $\lambda_m \approx \lambda_\pi$, the antiferromagnetic type order would change to a ferromagnetic type order with a non-zero polarization density $\vec{P}$ perpendicular to the 3-hypersphere.

The present day Universe expands with the acceleration of the order of $10^{-9} m/s^2$ and hence the virtual dipoles are subject to a very weak gravitational field. In the case of a strongly accelerated expansion with a very large acceleration, even dipoles with $\lambda_m \ll \lambda_\pi$ may be forced to align along the acceleration. At very large accelerations, particles more massive than pions may than as well contribute to the gravitational mass (dark energy) of the quantum vacuum.

## 5. Conclusion

In this Letter we have discussed how the gravitational properties of antimatter, together with the geometry of a hyper-spherical shell of thickness $\lambda_\pi$, might allow the identification of the observed dark energy density with the appropriate gravitational mass density of the quantum vacuum. In spite of some shortcomings and "derivation" that is not rigorous, the Eq. (4) provokes with its simplicity, fundamental appearance and accuracy.

In the framework of the Standard Model of elementary-particle physics, the vacuum energy density receives contributions from any quantum field. Hence, QCD fields provide just one (and not the largest) contribution to the vacuum energy. Contrary to the Standard Model, our approach, approach [2] and "coincidences" involving pions [3]-[5], suggest that only the QCD vacuum might somehow be important in Cosmology.

I would like to thank the Theoretical Physics Division at CERN where this work was performed. I am particularly grateful to Prof. John Ellis for his famous assistance to physicists coming from non-member States.

**Appendix**

Independently of the above consideration let us observe that Eqs. (3) and (4) can be related in an intriguing way. The Eq. (3) reduces to Eq. (4) applying a simple rule: use the mass of a pion as a cut-off on the right-hand side of Eq. (3) and multiply it by $2\lambda_\pi/R$.

The multiplication by $\lambda_\pi/R$ has an amusing interpretation. The Eq. (3) is based on the assumption that inertial and gravitational mass are equivalent; the source of gravitation are gravitational monopoles with a positive gravitational mass. But what if the sources of gravitation are gravitational dipoles? Let us compare the gravitational field produced by two positive monopoles (at mutual distance $\lambda_\pi$) and a dipole (i.e. a positive and a negative monopole also at distance $\lambda_\pi$). In full analogy with the electric dipole, the gravitational field produced by a dipole at large distance $R \gg \lambda_\pi$ is $\lambda_\pi/R$ weaker than the corresponding field produced by two monopoles.

The fact that the result (3), valid for the gravitational monopoles, must be multiplied by $2\lambda_\pi/R$ in order to obtain relation (4) suggests the existence of the gravitational dipoles; and the simplest candidate for gravitational dipoles are virtual particle-antiparticle pairs in the quantum vacuum.